\documentclass[prb,twocolumn,superscriptaddress,floatfix]{revtex4}
\usepackage{hyperref}
\usepackage{graphicx} 
\usepackage{amsmath}
\usepackage{float}
\usepackage[dvips]{epsfig}
\newcommand{\beq}{\begin{equation}}
\newcommand{\eeq}{\end{equation}} 
\newcommand{\beqa}{\begin{eqnarray}}
\newcommand{\eeqa}{\end{eqnarray}}
\newcommand{\ba}{\begin{array}}
\newcommand{\ea}{\end{array}}

\begin{document}

\title{Breakup of quantum liquid filaments into droplets}

\author{Francesco Ancilotto}
\affiliation{Dipartimento di Fisica e Astronomia ``Galileo Galilei''
and CNISM, Universit\`a di Padova, via Marzolo 8, 35122 Padova, Italy}
\affiliation{ CNR-IOM Democritos, via Bonomea, 265 - 34136 Trieste, Italy }

\author{Manuel Barranco}

\affiliation{Departament FQA, Facultat de F\'{\i}sica,
Universitat de Barcelona. Diagonal 645,
08028 Barcelona, Spain}
\affiliation{Institute of Nanoscience and Nanotechnology (IN2UB),
Universitat de Barcelona, Barcelona, Spain.}

\author{Mart\'{\i} Pi}
\affiliation{Departament FQA, Facultat de F\'{\i}sica,
Universitat de Barcelona. Diagonal 645,
08028 Barcelona, Spain}
\affiliation{Institute of Nanoscience and Nanotechnology (IN2UB),
Universitat de Barcelona, Barcelona, Spain.}

\begin{abstract} 
We have investigated how the Rayleigh-Plateau instability 
of a filament made of a $^{41}$K-$^{87}$Rb self-bound mixture may lead to an array
of identical quantum droplets, with typical breaking times which are shorter than the lifetime of the mixture.
If the filament is laterally confined --as it happens in a toroidal trap-- 
and atoms of one species are in excess with respect to the optimal, equilibrium ratio, 
the droplets are immersed into a superfluid  background made by the excess species which
provides global phase coherence to the system, suggesting that the droplets array in the unbalanced
system may display supersolid character. 
This possibility has been investigated by computing the 
non-classical translational inertia coefficient.
The filament may be a reasonable representation of a self-bound mixture subject to
toroidal confinement when the 
bigger circle radius of the torus is much larger than the filament radius.

\end{abstract} 
\date{\today}

\maketitle

\section{Introduction}

A new quantum state of matter has been predicted \cite{petrov2015quantum}
to occur in ultracold atomic gases composed of binary mixtures of Bose atoms where
the competition between the
inter-species attractive interactions
and quantum fluctuations, which produce a repulsive interaction,
may lead to the formation
of self-bound liquid droplets with ultralow
densities about eight orders of magnitude
lower than those, e.g., of the prototypical quantum fluid, namely liquid Helium.
Such novel quantum state was first observed experimentally 
in dipolar Bose gases \cite{Pfau,Ferlaino},
by exploiting the competition between contact repulsion and dipole-dipole attraction,
and later in
ultracold binary mixtures of Bose atoms
\cite{cabrera2018quantum,semeghini2018self,derrico2019observation,NaRb2021}.

Several binary Bose mixtures which may convert into quantum liquids 
have been investigated so far.
At variance with the 
largely studied short-lived homonuclear 
mixture of $^{39}$K atoms in two 
different hyperfine states,
the heteronuclear $^{41}$K-$^{87}$Rb mixture studied in Ref. \onlinecite{derrico2019observation}, 
which forms a quantum liquid state when 
the $^{41}$K-$^{87}$Rb scattering length $a_{12}$ becomes lower than the
critical value $a_{12}=-73.6\,a_0$ \cite{derrico2019observation}
($a_0$ being the Bohr radius), is rather long-lived with lifetimes
of the order of several tens of milliseconds,
i.e., more than an order of magnitude
larger than those
characterizing the $^{39}$K mixtures \cite{semeghini2018self}.
This opens the possibility of studying phenomena 
whose dynamical development requires time periods in the millisecond range.
One such phenomenon is the dynamical instability of quantum liquid filaments
leading to quantum drops formation, which is the subject of this work.

Liquid filaments 
(i.e. threads of liquid with the approximate shape of straight, long cylinders)
--and their dynamical instabilities-- are  
thoroughly studied subjects in classical fluids dynamics
both because of the underlying fundamental physical properties 
and  their potential applications. 
Experiments in this field, mainly concentrated on viscous fluids, 
are interpreted using theoretical approaches based on the solution of the
Navier-Stokes  equation subject to appropriate boundary conditions.
For an extended review on the subject see e.g. 
Ref. \onlinecite{Egg08} and references therein.

The stability of a macroscopic liquid filament, modelled with an 
infinitely extended
cylinder of radius $R$, was studied by Plateau,\cite{Pla57}
showing that it exists in an unstable equilibrium, and
any perturbation with wavelength $\lambda $ greater than $2\pi R$
triggers an instability where the surface 
tension breaks  the cylinder into droplets.
Lord Rayleigh later showed\cite{Ray79} that for an inviscid 
and incompressible liquid the 
fastest growing mode occurs when the wavelength
of the axial undulation that leads to the fragmentation
of the liquid filament into droplets is equal to  
$\lambda_c = 9.01\, R$  or, equivalently,
$kR=0.697$, where $k=2\pi /\lambda_c $ (Rayleigh-Plateau instability).
When the filament breaks up, one or more small satellite drops -resulting from 
the necks breaking- may form between the larger droplets.

Rayleigh-type instabilities are not limited to classical fluid only,
but may affect also quantum fluids. 
The dynamics of contraction and breaking 
of zero temperature superfluid $^4$He liquid thin
filaments in vacuum, triggered by the above kind of instabilities,
has recently been addressed\cite{Anc23}
using a $^4$He Density Functional Theory approach which
accurately describes superfluid $^4$He at zero temperature\cite{Anc17}.

We investigate here the instability of thin filaments made of 
another superfluid system, namely a quantum liquid made of
an ultracold bosonic  $^{41}$K-$^{87}$Rb  mixture. 
We notice that the 
instability of a two-component Bose-Einstein condensate
has been investigated \cite{saito} in the repulsive (immiscible) regime, where 
a cylindrical condensate made of one species surrounded by the other component was found to
undergo breakup into gaseous bubbles.

We consider here a linear, thin filament 
with periodic boundary conditions imposed at its ends.
One can consider such geometry as a limiting case 
of a mixture subject to a toroidal confinement when the 
bigger circle radius of the torus is much larger than the filament radius.

To study the instability of quantum liquid filaments we will use two 
different approaches. 
One is the widely known Mean-Field+Lee-Huang-Yang 
(MF-LHY) approximation which provides a reliable description 
of the binary mixture in the quantum liquid regime through the
solution in three dimensions of two coupled non-linear Gross-Pitaevskii (GP) equations,
and which applies to arbitrary concentrations of the two species.
Since in the quantum liquid state of the $^{41}$K-$^{87}$Rb uniform mixture 
the equilibrium densities of the two components 
are expected to have a fixed ratio, the system 
can be effectively described by one single wave function
satisfying an effective GP equation\cite{Pol21}.
Inspired in the work carried out in Ref. \onlinecite{Sal02} for one-component 
BEC gases, in a second approach
we will use a variational formalism suited to the geometry 
we are implementing in the present work. This allows one to reduce the coupled three-dimensional (3D) GP equations 
to one effective one-dimensional (1D) GP equation plus an algebraic equation.
We verify the feasibility of this simplified approach by comparing the filament properties it yields 
with the ones obtained by solving the 3D coupled GP equations. Either approach will
disclose the time scale for filament breakup and the appearance of quantum droplets
as a result of its fragmentation.

In the second part of this work we will address an interesting aspect arising when    
one of the species is in excess with respect to the optimal density ratio and
the droplets resulting from filament breakup are 
immersed in a superfluid background made by the species in
excess, which provides global phase coherence to the system
and may lead to supersolid behavior.
Such possibility is investigated by computing the 
non-classical translational inertia associated to this system.

This work is organized as follows. In Sec. II we  review the theoretical approach used
to describe the binary mixture in the quantum liquid regime, which is based on the MF-LHY approximation,
and also introduce a simpler yet accurate variational approach 
based on a 1D effective equation introduced some time ago 
to describe single-component Bose Einstein condensates subject to tight 
radial harmonic confinement \cite{Sal02}.  Such approach is extended here to the 
quantum liquid mixture case and is used to address the dynamical instability 
of quantum liquid filaments leading to quantum droplets formation.
The results are presented in Sec. III, and a summary is given in Sec. IV.

\section{Method}

The Gross-Pitaevskii  energy functional for a Bose-Bose mixture,
including the Lee-Huang-Yang 
correction accounting for quantum fluctuations beyond mean-field
reads \cite{petrov2015quantum,ancilotto2018self}

\begin{widetext}
\begin{equation}
E = \sum_{i=1}^{2}\int d{\bf r} \, \left[{\hbar^2 \over 2m_i}|\nabla \psi_{i}({\bf r})|^2  
+ V_{i}({\bf r}) \rho_{i}({\bf r})\right]\,
+\frac{1}{2}\sum_{i,j=1}^{2}g_{ij}\int d{\bf r} \, \rho_{i}({\bf r})\rho_{j}({\bf r}) \,
+\int d{\bf r} \, {\cal E} _{\rm LHY}(\rho_1({\bf r}),\rho_2({\bf r}))\, ,
\label{eq:energyfun}
\end{equation}
\end{widetext}
where $V_i({\bf r})$ and $\rho_i({\bf r})=|\psi _i({\bf r})|^2$ 
represent the external potential and the number density of each component
($i=1$ for $^{41}$K, $i=2$ for $^{87}$Rb), respectively. The coupling constants are
$g_{11}=4\pi  a_{11} \hbar^2/m_1$, $g_{22}=4\pi a_{22} \hbar^2/m_2$, and $g_{12}=g_{21}=2\pi a_{12} \hbar^2/m_r$,
where $m_r =m_1m_2/(m_1+m_2)$ is the reduced mass.
The number densities $\rho_1,\,\rho_2$ are normalized such
that $\int _V \rho_1({\bf r})\,{\rm d}{\bf r} =N_1$ and $\int _V \rho_2({\bf r})\,{\rm d}{\bf r} =N_2$.
The intra-species $s$-wave scattering lengths $a_{11}$ and $a_{22}$ are both positive, while the inter-species
one, $a_{12}$, is negative.
The scattering parameters describing the intraspecies repulsion
are fixed and their values are equal to $a_{11} = 65\,a_0$ \cite{derrico2007feshbach} 
and $a_{22} = 100.4\,a_0$ \cite{marte}. 
The heteronuclear scattering length $a_{12}$ can be tuned 
by means of Feshbach resonances. The onset of the mean-field collapse regime 
leading to the quantum liquid state corresponds to $g_{12}+\sqrt{g_{11}g_{22}}=0$, 
which occurs at $a_{12}=-73.6\,a_0$ \cite{derrico2019observation}.

The LHY correction is \cite{petrov2015quantum,ancilotto2018self}
\begin{align}
{\cal E} _{\rm LHY} &= {8\over 15 \pi^2} \left(\frac{m_1}{\hbar^2}\right)^{3/2}
\!\!\!\!\!\!(g_{11} \rho_1)^{5/2}
f\left(\frac{m_2}{m_1},\frac{g_{12}^2}{g_{11}g_{22}},\frac{g_{22}\,\rho_2}{g_{11}\,\rho_1}\right)
\nonumber
\\
&\equiv \mathcal{C} (g_{11}\rho_1)^{5/2}f(z,u,x).
\label{functional}
\end{align}
Here, $f(z,u,x)>0$ is a dimensionless function whose explicit expression
for an heteronuclear mixture 
can be found in Ref. \onlinecite{ancilotto2018self}.
Following Ref. \onlinecite{petrov2015quantum}, we consider this
function at the mean-field collapse $u=1$, i.e., ${\cal E} _{\rm LHY}=\mathcal{C} (g_{11}\rho_1)^{5/2}f(z,1,x)$.
We note that the actual expression for $f$ can be fitted very accurately with the
same functional form of the homonuclear  ($m_{1}=m_{2}$) case\cite{Minardi}
\begin{equation}
f\left(z,1,x\right) = \left(1+z^{\alpha_1} x\right)^{\beta_1} \, ,
\label{flhy}
\end{equation}
where $\alpha_1$ and $\beta_1$ are fitting parameters.
For the $^{41}$K-$^{87}$Rb mixture ($z=87/41$) one has \cite{Minardi} $\alpha_1 = 0.586$ and $\beta_1= 2.506$.
We will use the form in Eq. \eqref{flhy} for our calculations.

\subsection{3D equations}

Minimization of the action associated to Eq.~(\ref{eq:energyfun}) leads to
the following Euler-Lagrange equations (\textit{generalized} GP equations) 
\begin{equation}
i \hbar {\partial \psi _i \over \partial t} =
\left[-{\hbar^2 \over 2m_i}\nabla ^2 + V_i + \mu_{i}(\rho_1,\rho_2) \right]\psi _i 
 \, ,
\label{eq:gpe}
\end{equation}
where
\begin{equation}
\mu_{i} = g_{ii}\rho_i+g_{ij}\rho_j+
\frac{\partial {\cal E}_{\rm LHY}}{\partial \rho_i}\quad \,( j \ne i) 
\label{eq:chempot}
\end{equation}
and
\begin{align}
\frac{\partial {\cal E} _{\rm LHY}}{\partial \rho_1} &=
\mathcal{C} \,g_{11}(g_{11}\rho_{1})^{3/2}\left({5\over 2}f-x{\partial f\over \partial x}\right)
\\
\frac{\partial {\cal E} _{\rm LHY}}{\partial \rho_2} &=
\mathcal{C} \,g_{22}(g_{11}\rho_{1})^{3/2}{\partial f\over \partial x} \, ,
\end{align}
where $\mathcal{C}$ is defined in Eq. \eqref{functional}.

The numerical solutions of Eqs. \eqref{eq:gpe} provide 
the time-evolution of a $^{41}$K-$^{87}$Rb mixture with arbitrary compositions $N_1,N_2$ 
in three-dimensions. In the following, we will refer to this solution as the 3D model  to 
distinguish it from a simpler, computationally faster 1D model approach  which we will describe next.

\subsection{Effective 1D equation}

The mixture of the two bosonic species in the homogeneous phase 
is stable against fluctuations in the 
concentration $N_1/N_2$ if \cite{petrov2015quantum}
\begin{equation}
\dfrac{\rho_1}{\rho_2} = \sqrt{\dfrac{g_{22}}{g_{11}}} \ .
\end{equation}
For the $^{41}$K-$^{87}$Rb system investigated in the present work, $\rho_1 / \rho_2=0.853$.
As pointed out in Refs. \onlinecite{petrov2015quantum,ancilotto2018self, 
staudinger2018self}, it is safe to assume that this optimal composition is 
realized everywhere in the system. 
Therefore, the energy functional Eq. \eqref{eq:energyfun}
becomes effectively single-component and can be expressed
in terms of the density $\rho_1$ alone as $E=\int \mathcal{E}(\mathbf{r})$ d$\mathbf{r}$, where
\begin{equation} \label{en_dens}
\mathcal{E} = \alpha\frac{(\nabla \rho_1)^2}{\rho_1} + \beta \rho_1^2 + \gamma \rho_1^{5/2}
\end{equation}
with 
\begin{equation}\label{alpha}
        \alpha  = \frac{1}{4}\left( 
\frac{\hbar^2}{2m_1}+\frac{\hbar^2}{2m_2}\sqrt{\frac{g_{11}}{g_{22}}}\right) 
        \end{equation}
        \begin{equation}\label{beta}
        \beta = g_{11}+ g_{12}\sqrt{\frac{g_{11}}{g_{22}}} 
        \end{equation}
        \begin{equation}\label{gamma}
        \gamma =  {\cal C} \, g_{11}^{5/2} f(z,1,x) \, .
\end{equation}
The 3D differential equation governing the evolution of the
macroscopic wave function $\Psi (\mathbf{r},t)$ of the system,
such that $|\Psi |^2=\rho _1$, is
\begin{equation}
\left[-\frac {\hbar ^2}{2m} \nabla ^2 + V_{ext}(\mathbf{r}) + \frac {\partial \mathcal{E}}{\partial \rho_1}
\right]\Psi (\mathbf{r},t) =i\hbar \frac {\partial \Psi (\mathbf{r},t)}{\partial t} \, ,
\label{3d_gpe}
\end{equation}
where $V_{ext}$ is any external potential acting on the system and 
\begin{equation}
m=\left[\frac{1}{m_1} +\frac {1}{m_2}\sqrt{\frac {g_{11}}{g_{22}}}  \right]^{-1} \, .
\label{eff_mass}
\end{equation}

Equation (\ref{3d_gpe}) can be derived by applying the quantum least action principle 
to the action
\begin{equation}
S=\int dt \int d\mathbf{r}\,\Psi ^\ast (\mathbf{r},t)\left[ i\hbar \frac {\partial }{\partial t} -\hat{H} \right]\Psi (\mathbf{r},t) \, ,
\label{action}
\end{equation}
where 
\begin{equation}
\hat{H}=-\frac {\hbar ^2}{2m} \nabla ^2 + V_{ext} + \frac {{\cal E}}{\rho_1} \, ,
\label{hamilt}
\end{equation}
${\cal E}/\rho_1$ being the energy per particle, $E/N_1$, in the homogeneous system
\begin{equation}
\frac {\mathcal{E}}{\rho _1}=\beta \rho_1 + \gamma \rho_1^{3/2} \, .
\end{equation}
The external potential is taken here 
in the form of harmonic confinement in the transverse direction 
(in the $x-y$ plane) and
generic in the axial ($z$) direction:
\begin{equation}
V_{ext}=\frac {1}{2}m_1 \omega _\perp ^2\,(x^2+y^2)\,+\,V(z)
\end{equation}
We assume, as often done in the experiments on heteronuclear mixtures, that 
$m_1\omega _1^2=m_2\omega_2 ^2$, $\omega _1$ and $\omega _2$ being the 
frequencies of the harmonic confinements acting on the two species,
$V_i=(1/2)m_i\omega_i ^2 \,(x^2+y^2)$ in Eq. \eqref{eq:energyfun}.
Therefore, $\omega _\perp ^2 =\omega_1 ^2 (1+\sqrt{g_{11}/g_{22}})$.

We follow the approach of Ref. \onlinecite{Sal02}, where an effective 1D wave equation
can be derived using a variational approach which describes the axial dynamics of
a Bose-Einstein condensate confined in an external potential with cylindrical symmetry
around the $z$-axis.
The action functional Eq. \eqref{action} is minimized using the following trial wave function
\begin{equation}
\Psi (\mathbf{r},t) =\varphi(x,y,t;\sigma (z,t))\,h(z,t) \, ,
\label{varpsi}
\end{equation}
where the transverse part of the wave function is modelled by a Gaussian
\begin{equation}
\varphi(x,y,t;\sigma (z,t))=\frac {1}{\pi ^{1/2}\sigma(z,t)}e^{-(x^2+y^2)/2\sigma^2(z,t)} \, .
\end{equation}
While $\varphi $ is normalized to unity, $h$ is normalized according 
to the number of atoms in the species 1, $\int |h|^2 dz=N_1$.
We will show in the following that, for not too high
atomic densities, the choice of a Gaussian
to describe the wave function in the transverse plane is indeed
appropriate for the investigated system.
 
The variational functions $\sigma(z,t)$ and $h(z,t)$ are determined
by minimizing the action functional after integrating in the $(x,y)$ plane.
A further assumption is made in Ref. \onlinecite{Sal02}, namely
that the transverse wavefunction is slowly varying along the axial direction,
meaning that  $\nabla ^2 \varphi\sim \nabla _\perp ^2\varphi $, where $\nabla _\perp ^2 =\partial ^2/\partial x^2+
\partial ^2/\partial y^2$. 
After inserting Eq. \eqref{varpsi} into the action and  integrating in the $(x,y)$ plane,
the action functional becomes

\begin{widetext}
\begin{equation}
S=\int dt \int dz \; h^\ast \left[ i\hbar \frac {\partial }{\partial t} +\frac {\hbar ^2}{2m} 
\frac {\partial ^2}{\partial z^2}-V(z)-\frac {\hbar ^2}{2m}\sigma ^{-2}-\frac {1}{2}m\omega _\perp ^2 
\sigma ^2-\frac {\beta |h|^2}{2\pi }\sigma ^{-2} -\frac {2\gamma |h|^3}{5\pi^{3/2}}\sigma ^{-3}\right]h \, .
\label{action_1}
\end{equation}
\end{widetext}

From the variational minimization 
of the above functional, $\delta S/\delta h^\ast =0$ and $\delta S/\delta \sigma =0$, one obtains
the following two equations

\begin{widetext}
\begin{equation}
i\hbar \frac {\partial h}{\partial t}=
\left[-\frac {\hbar ^2}{2m}\frac {\partial ^2}{\partial z^2}+V(z) 
+\frac {\beta |h|^2}{\pi \sigma ^2} 
+\frac {\gamma \sigma ^{-3}}{\pi^{3/2}}|h|^3 
+\frac {\hbar ^2}{2m}\sigma ^{-2} +\frac {1}{2}m \omega _\perp ^2 \sigma ^2\right]h =0
\label{el_2}
\end{equation}
\end{widetext}

\begin{equation}
\frac {\hbar ^2}{2m}\sigma ^{-3} -\frac {1}{2}m \omega _\perp ^2 \sigma 
+\frac {\beta |h|^2}{2\pi} \sigma ^{-3}
+\frac {3\gamma |h|^3}{5 \pi^{3/2}}\sigma ^{-4}=0 \, .
\label{el_1}
\end{equation}
Notice that the case of a single-species Bose-Einstein condensate confined in an external potential
with axial symmetry is recovered when $\gamma =0$ and $\beta =g/2$, $g$ 
being the scattering amplitude of the contact atom-atom interaction. In this 
case, the equations above reduce to Eqs. (6) and (7)
of Ref. \onlinecite{Sal02}.
The main advantage of this formulation is that 
the computational cost of finding the time-evolution of the system 
is low, being reduced essentially to the numerical solution of a non-linear
1D Schr\"odinger equation instead of solving the more demanding 3D 
Eqs. \eqref{eq:gpe} or \eqref{3d_gpe}.

The above equations are solved by propagating the wave function $h(z,t)$
in imaginary time, if stationary states are sought, or 
in real time to simulate the dynamics of the system
starting from specified initial states.
In both cases, the system
wave function 
is mapped onto an  
equally spaced 1D Cartesian grid and the differential operator 
is represented by a 13-point formula.
At each time step, Eq. \eqref{el_1}  
is solved using a simple bisection method
to provide an updated value for the function $\sigma(z,t)$ to be used in
Eq. \eqref{el_2}. The latter is solved in real time
by using  Hamming's predictor-modifier-corrector method 
initiated by a fourth-order
Runge-Kutta-Gill algorithm \cite{Ral60}. 
Periodic boundary conditions (PBC) are imposed along the $z$-axis.
The spatial mesh spacing and time step are chosen such that,
during the real time evolution, excellent conservation
of the total energy of the system is guaranteed.

\section{Results}

\subsection{Time scales for breakup}

Possible experimental observations of the instabilities described in this work
will only be possible if the characteristic time for breakup of a thin  $^{41}$K-$^{87}$Rb
filament is smaller that the lifetime of the mixture.
The time scale for instability and breakup
of a liquid filament in the form of a cylinder with radius $R$ 
made of an incompressible fluid with bulk number density $\rho_0$
and surface tension $\mathcal{T}_0$ is set
by the capillary time $\tau_c$ defined as
$ \tau_c = \sqrt{m \rho_0 R^3/\mathcal{T}_0}$, $m$ being the mass of the atoms
in the liquid\cite{Egg08}.
In the case of a binary mixture we generalize this definition as
\begin{equation}
\tau_c = \sqrt{\frac{(m_1\rho_1+m_2\rho_2) R^3}{\mathcal{T}_0 }} \, .
\label{tau_cap}
\end{equation}
The surface tension of the binary mixture $^{41}$K-$^{87}$Rb has been computed
for different values of the interspecies scattering length $a_{12}$ in Ref. \onlinecite{Pol21}.
It turns out that relatively
small changes in the interspecies interaction strength cause
order-of-magnitude changes in the surface tension\cite{Pol21}, which ranges
from $\mathcal{T}_0 \sim 10^2\,nK/\mu m^2$ for $a_{12}=-80\,a_0$
to $\mathcal{T}_0 \sim 10^5\,nK/\mu m^2$ for $a_{12}=-100\,a_0$.

We must notice that 
the surface tension values quoted above
have been obtained for a planar interface 
though in finite systems like quantum liquid filaments or droplets there is a contribution due
to the interfacial curvature of the surface. The
curvature-dependent surface tension can be expressed in terms
of the so-called Tolman length \cite{Tol49} $\delta $. To  first approximation,
the Tolman length can be taken independent of the droplet size and 
 the size-dependent surface tension can be expressed in terms of that of the
planar surface as \cite{Blo06}
\begin{equation}
\mathcal{T}(R)=\mathcal{T}_0\,\left(1-\frac {2\delta }{R}\right)  \, ,
\end{equation}
where $R$ is the radius of curvature.
The Tolman length for the $^{41}$K-$^{87}$Rb 
binary mixture in the self-bound state 
has been computed in the MF-LHY approach \cite{Pol21}.
For the case studied here, i.e. $a_{12}=-90\,a_0$, it has been found that
$\delta =-2.95\times 10^3\,a_0$, making the curvature-dependent surface tension larger than the 
planar surface one. For typical radii $R\sim 2\times 10^4\,a_0$ (see the following), 
the correction amounts to a sizeable $2\delta /R\sim 0.1$. 

\begin{figure}
\centerline{
\includegraphics[width=1.0\linewidth,clip]{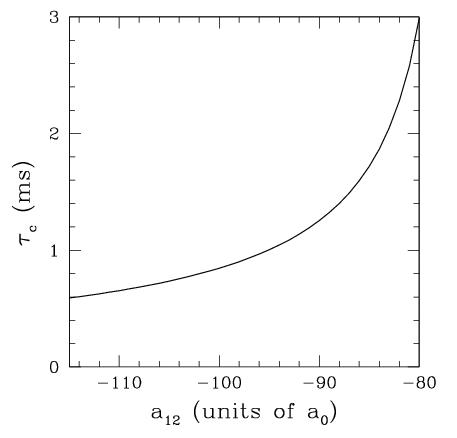}
}
\caption{Capillary time $\tau_c$ as a function of $a_{12}$ 
calculated for a $^{41}$K-$^{87}$Rb cylindrical filament with $R=15,000\,a_0$
}
\label{fig1}
\end{figure}

The capillary time calculated by Eq. \eqref{tau_cap} 
using the planar value $\mathcal{T}_0$ is plotted in Fig. \ref{fig1} as 
a function of the interspecies scattering length $a_{12}$ taking $R=15000\,a_0$
as a value representative of the typical sized investigated here.
It appears that, even for values of $a_{12}$ close to the onset of the self-bound regime
($a_{12}\sim -74\,a_0$), the capillary time is much smaller than the typical lifetime
of the $^{41}$K-$^{87}$Rb mixture in the quantum liquid regime (several tens of milliseconds)\cite{derrico2019observation}.
Notice however that the actual time taken for the filament to break 
into droplets depends upon the amplitude
of the initial density perturbation triggering the instability.
This will be discussed in Sec. III.C.

\subsection{Equilibrium structure of a free-standing cylindrical filament}

Prior to the dynamics, one has to obtain the static configuration constituting its starting point.  
To this end, we have computed
the equilibrium structure of a free-standing cylindrical filament,
i.e. $V(z)=0$ and $\omega_\perp =0$ in Eq. \eqref{el_2}, taking 
$a_{12}=-90 \,a_0$. 

Some properties of the equilibrium filament are reported in the Table \ref{table1}. They have been
computed using the 1D equation for three values of the linear density $N_1/L$, where $L$ is
the length of the filament, which yield three different radial density profiles and sizes.
In the table, the sharp radius $R$ of the cylinder is defined as the 
radial distance at which
$\rho=\rho_c/2$, $\rho_c=N_1/L$ being the density along the filament axis.
Notice that this definition is strictly appropriate in the case of a
thick filament consisting in a bulk region with flat-top density equal to the 
equilibrium density in the homogeneous system separated from 
the vacuum by a finite-width surface profile, the surface width being much smaller 
than $R$. In the present case, this definition of radius is
somewhat arbitrary due to the Gaussian-like nature of the transverse density profile,
as shown in the following.
The capillary times have been computed from Eq. \eqref{tau_cap} using for the filament radius  either 
the sharp radius, or the equilibrium value for $\sigma$.
The definition and values of the Rayleigh-Plateau instability length $\lambda _c$ in Table \ref{table1} are 
discussed in  Sec. III.C.
 
\begin{table*}
\begin{tabular}{lcccccccccc}
\hline
& $N_1/L $ ($a_0^{-1}$) & $E/(N_1+N_2)$ $ (Ha)$ &  $\sigma $ $(a_0)$  & $R$ $(a_0)$ & $\tau_c^{\sigma }$ (ms) 
& $\tau_c^{R}$ (ms) & $\lambda_c $ $(a_0)$ & $2\pi R/\lambda_c$ \\
\hline
 & 0.095  &  -1.295 $10^{-14}$ & 16045 & 13368 & 1.101 & 0.837   &  72910 & 1.15\\
 & 0.159  & -2.189 $10^{-14}$ & 17390 & 14487 & 1.243  & 0.945  &  111300 & 0.82\\
 & 0.238 & -2.759 $10^{-14}$ & 19715 & 16423 & 1.500 & 1.141   &   160680 & 0.64 \\
\hline
 \end{tabular}
  \caption{Equilibrium properties for three filaments with $a_{12}=-90\,a_0$ 
calculated with the 1D equation. Atomic units are used for lengths and energies, times are expressed in ms.
 \label{table1}
}
\end{table*}

\begin{figure}
\centerline{
\includegraphics[width=1.0\linewidth,clip]{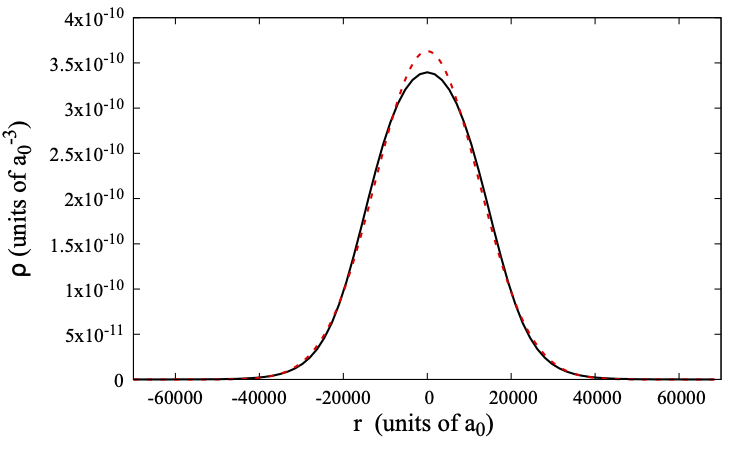}
}
\caption{Density profile for the filament with $N_1/L=0.159\,a_0^{-1}$ plotted 
in the transverse direction (perpendicular to the filament axis).
The total density $\rho_1+\rho_2$ is shown in units of $a_0^{-3}$.
Solid line: 3D model; dashed line: 1D model.
}
\label{fig2}
\end{figure}

To verify the validity of the approximations underlying the use of 
Eqs. \eqref{el_2} and \eqref{el_1}, we have also computed the 
equilibrium structure for the same cylindrical filament using instead the
3D Eqs. \eqref{eq:gpe}.
The transverse profile of the filament computed with the two methods
(1D and 3D equations) for $N_1/L=0.159\, a_0^{-1}$ is shown in Fig. \ref{fig2}, where one may see a good
agreement between both and conclude that it is fairly Gaussian-like.
Notice  that the density profile of the filament 
is very different from that of a macroscopic
incompressible liquid cylinder 
characterized by a flat-top density profile encompassing
a bulk region with a nearly constant
density, and a narrow surface region whose width is determined by
the surface tension \cite{Pol21}, as those found for classical viscid 
fluids\cite{Egg08} and superfluid $^4$He\cite{Anc23} filaments.
Here, we have instead an all-surface, highly compressible
cylindrical filament. Experimentally, it is easier to realize this Gaussian-like
system, since the very large number of atoms required to 
create a flat-top density profile is difficult to reach due to the increasing
role played by three-body losses which rapidly deplete the system.

\begin{table*}
\begin{tabular}{lcccccccccc}
\hline
& $N_1/L $ ($a_0^{-1}$) & $E/(N_1+N_2)$ $ (Ha)$ &  $\sigma $ $(a_0)$  & $R$ $(a_0)$ & $\tau_c^{\sigma }$ (ms) & $\tau_c^{R}$ (ms)  
& $\lambda_c $ $(a_0)$ & $2\pi R/\lambda_c $ \\
\hline
 & 0.095  & -1.333 $10^{-14}$  & 15769  & 12951 & 1.101 & 0.837 & 73700 & 1.10   &  &\\
 & 0.159  & -2.214 $10^{-14}$ &  17917 & 15650 & 1.243 & 0.945 & 103560 & 0.95  &  &  \\
 & 0.238  & -2.815 $10^{-14}$ &  21019 & 19178 & 1.500 & 1.141 & 135200 & 0.89  &  &  \\
\hline
 \end{tabular}
  \caption{Equilibrium properties for three filaments with $a_{12}=-90\,a_0$ 
calculated with the 3D equations. Atomic units are used for lengths and energies, times are 
expressed in ms.
\label{table2}
}
\end{table*}

We report in Table \ref{table2} some equilibrium properties calculated using the 3D equations for the same filaments as in Table \ref{table1}. 
The width $\sigma$ has been estimated by assuming that the
transverse wave function is a Gaussian as in Eq. (\ref{varpsi}), 
i.e. $\sigma =\sqrt{N_1/(\pi L \rho_c)}$,
where 
$\rho_c $ is the uniform density
value along the filament axis. 
A comparison with Table \ref{table1}
shows an overall good agreement 
between 1D and 3D models,
albeit with some differences in the case of the thicker filament.

\subsection{Capillary instability}
\label{section}

We have verified by real-time dynamics that a cylindrical quantum liquid 
filament is indeed unstable against a small initial 
axial perturbation of the density with a sufficiently large wavelength, as predicted
by the Rayleigh theory.
We have used 
both approaches, 
i.e., that based 
on the 3D MF-LHY equations and the one based on the effective 1D equation.
For a classical incompressible fluid, 
any perturbation with wavelength $\lambda $ greater than $2\pi R$
makes the system unstable allowing the surface tension 
to break  the cylinder into droplets, thus
decreasing the surface energy of the system.

We have studied the instability threshold for the three filaments 
whose properties are summarized in Tables \ref{table1} and \ref{table2}. 
To this end, we have applied a weak axial perturbation on the transverse width 
of the filament of the form
\begin{equation}
\sigma (z)=\sigma_0\left[1-\epsilon \,cos\left(\frac{2\pi z}{L}\right)\right] \, ,
\end{equation}
where $L$ is the total length of the filament,
$\sigma _0$ is the equilibrium value for the filament transverse size,
and $\epsilon \ll 1$.
We then let the system evolve in time; if $L$ is smaller 
than a critical value $\lambda _c $, the filament remains
intact and one simply observes small amplitude surface oscillations due to the 
initial perturbation. However, above this critical value, 
the amplitude of the initial perturbation starts to grow,
a neck develops and eventually the filament breaks into droplets.
We determined in this way the value of $k=2\pi/L$ that makes
the filaments unstable,
i.e. the critical wavelength $\lambda _c=2\pi/k$ (see Tables \ref{table1}-\ref{table2}).
We recall that linear theory for a classical fluid filament predicts
$(2\pi/\lambda _c)R=1$ \cite{Ray79};
this is only approximately true for the liquid filaments 
investigated here,
where the ratio $R/\lambda _c$ turns out  not to be universal but weakly depends on the
linear density of the system or, equivalently, on the radius, as shown in Tables \ref{table1}-\ref{table2}.
We remark that this is not a limitation due to the nanoscopic nature of our
system; in fact, calculations on nanoscopic $^4$He superfluid liquid filaments \cite{Anc23}
yielded the universal result of linear theory for inviscid classical fluids. Rather, it
is likely due to the all-surface 
nature of the filaments investigated here, and the very large compressibility of
quantum liquids \cite{Pol21};
the classical, linear theory assumes instead an incompressible fluid and a sharp surface filament.

\begin{figure}
\centerline{
\includegraphics[width=1.0\linewidth,clip]{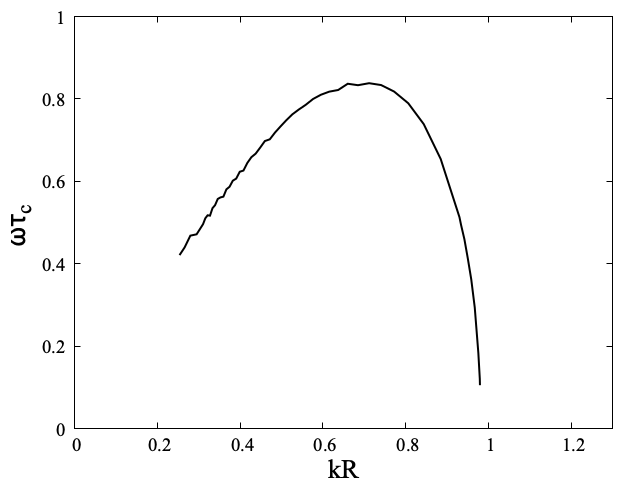}
}
\caption{Plot of $\omega \tau_c$ vs. $kR$ for the
filament with $N_1/L=0.159\,a_0^{-1}$. 
}
\label{fig3}
\end{figure}

The actual time taken for the filament to break into droplets depends upon the amplitude
of the initial density perturbation. It is defined as the time $\tau_b$
it takes for the wave amplitude with the largest frequency to grow up to 
the value of the filament radius, thus breaking it \cite{Dum08,Hoe13}.
For a given time, the amplitude of a perturbation, with some given 
wavevector $k$, evolves as
\begin{equation}
\delta (t)=\delta _0 e^{\omega (k) t} \, .
\end{equation}
We have computed the dynamics of neck shrinking by monitoring during the
real-time evolution the quantity
$\delta(t)=(R_{max}-R_{min})/2$, where
the radii $R_{max}$ and $R_{min}$ are measured at the two positions
corresponding to a crest (maximum) and a valley (minimum) in the
filament surface.
As for $^4$He filaments\cite{Anc23},
we have first checked that the exponential law is indeed strictly followed by 
the simulations, and from 
the calculated values for $\delta(t)/\delta_0$ we have
computed $\omega $ as a function of the adimensional quantity $kR$.
The results are shown in Fig. \ref{fig3} for 
the filament with linear density $N_1/L=0.159\,a_0^{-1}$.
It appears that there is a maximum frequency $\omega _{max}=0.84/\tau_c $ at about $kR=0.7$,
i.e., the actual breaking of a filament subject to a most
general perturbation will be dominated by the fastest mode with $\omega = \omega_{max}$
being characterized by a time constant $\tau \sim 2\pi/\omega_{max}\sim 9.3$ ms.

The actual dynamics of the instability is provided by the solutions
of the time dependent equation for the filament. We use here the 
3D equations described before and apply them to the  
$N_1/L=0.159\,a_0^{-1}$ free-standing filament, and to a filament with the same $N_1/L$ value 
laterally confined by a harmonic confinement with $\omega_x=\omega_y=2\pi (100)$ Hz.
We chose a value for the wavelength 
corresponding to the maximum of the $\omega (k)$ curve, i.e., $kR\sim 0.7$. 
Similarly to what we have done when solving the 1D problem, we 
started the numerical simulations from the previously obtained equilibrium filament and apply a small
axial perturbation with wavelength $\lambda _c$ and initial amplitude $\epsilon =0.05 R$.

\begin{figure}
\centerline{
\includegraphics[width=1.0\linewidth,clip]{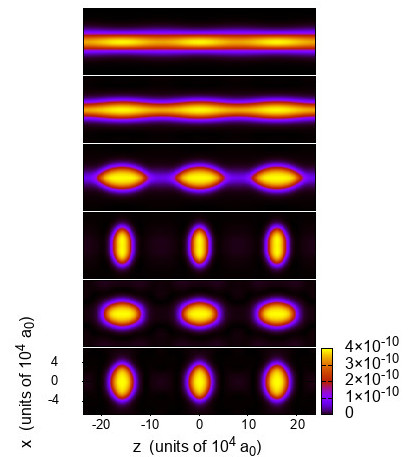}
}
\caption{ 
Snapshots from the real time evolution obtained by solving  the 3D equations for 
a free-standing filament with $N_1/L=0.159\,a_0^{-1}$. 
The color bar displays the total number density in units of $a_0^{-3}$.
The filament is initially perturbed by a periodic perturbation  with $kR\sim 0.7$.
From top to bottom, the snapshots are taken at $t= 0,15,19,21,24$, and 27 ms.
}
\label{fig4}
\end{figure}

Figure \ref{fig4} shows  some snapshots of the filament density on a symmetry plane containing the symmetry $z$-axis. It corresponds 
to the evolution of the free-standing filament.
It can be seen from the figure that, starting from the perturbed
filament, undulations whose amplitude increases with time appear 
along the filament. The instability is caused by the 
Laplace pressure increase in constricted regions (necks), driving 
out the fluid and hence reducing further the neck radius.
The filament evolves into higher density bulges connected by thin threads
bridging adjacent bulges.
At variance with the fragmentation of inviscid classical fluids and even 
superfluid $^4$He, where 
such threads eventually break up forming smaller satellite droplets,
here instead they swiftly evaporate.
The main droplets forming after the fragmentation 
execute oscillatory motion, being alternately compressed
and elongated in the filament direction.

\begin{figure}
\centerline{
\includegraphics[width=1.0\linewidth,clip]{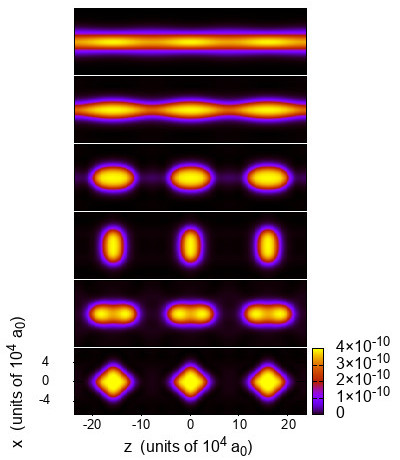}
}
\caption{
Snapshots from the real time evolution obtained by solving  the 3D equations for 
a filament with $N_1/L=0.159\,a_0^{-1}$ subject to a lateral harmonic confinement  
as described in the text. 
The color bar displays the total number density in units of $a_0^{-3}$.
The filament is initially perturbed by a periodic perturbation  with $kR\sim 0.7$.
From top to bottom, the snapshots are taken at $t=0,15,19,21,24,$ and 27 ms.
}
\label{fig5}
\end{figure}

In order to check that the fragmentation dynamics of the filament is 
not hindered by the presence of an 
external potential,
like the one necessary to confine the filament within a toroidal geometry,
we also addressed the case where the filament is subject to 
a transverse harmonic confinement.
This is shown in 
Fig. \ref{fig5} for the same instants shown in Fig. \ref{fig4}.
It appears that the dynamics of fragmentation is very similar to the case 
of the free-standing filament, with some visible effects of the lateral confinement 
on the droplet shapes (compare the last three panels in Fig. \ref{fig4} and Fig.\ref{fig5}).

\bigskip
\bigskip

We have also studied the instability of thin $^{41}$K-$^{87}$Rb free-standing filaments triggered by
a more general initial axial perturbation. 
To do so, we
consider a ``random'' modulation of the initial 
transverse section $\sigma (z)$ obtained by 
superimposing two sinusoidal modulations with incommensurate periods, one with
a short period $d_s$, and another with a longer period
$d_l$  (see, e.g., Ref. \onlinecite{Fal07}):
\begin{equation}
\sigma(z)=\sigma _0\,\{1+\epsilon [sin(\pi z/d_s)^2+sin(\pi z/d_l)^2-1]\} \, ,
\end{equation}
where $\sigma _0$ is the width of the equilibrium filament.

An infinite quasi-periodic disorder results when the ratio $d_l/d_s$ 
is an irrational number. We choose the golden ratio $\phi = 
(\sqrt{5} + 1)/2$ for such a number. Since our simulations use a
finite box with periodic boundary conditions,
to make the above expression of $\sigma $ consistent with 
the use of PBC along the $z$-axis one must
approximate this number by the ratio of two integer numbers, the largest one providing the total length
of the periodic cell used in the calculation. 
Here we approximate $\phi $ with the ratio of two successive numbers in
the Fibonacci sequence, $d_l/d_s = F_{j+1} /F_j$ \cite{Mod09}, which
notoriously converges towards the golden ratio for large values of $j$.
In particular, we take $L=8\,\lambda _c$ where 
$\lambda _c$ is the instability threshold for filament breaking.
We consider  the case $N_1/L=0.159\,a_0^{-1}$, so that $\lambda _c=111300$
(see Table I).
We choose the two adjacent elements $13,21$ in the Fibonacci sequence, 
 so that $d_s=42400\,a_0$ and $d_l=68492\,a_0$.
With this choice, $L=21\,d_s=13\,d_l$, 
making the applied 
perturbation satisfying the periodic boundary conditions.
We also took $\epsilon =0.06 R$.
Given the large size of the system, which makes 
the full 3D calculations very computationally demanding,
we have employed here the 1D effective equation. 

\begin{figure*}[!]
\centerline{
\includegraphics[width=1.0\linewidth,clip]{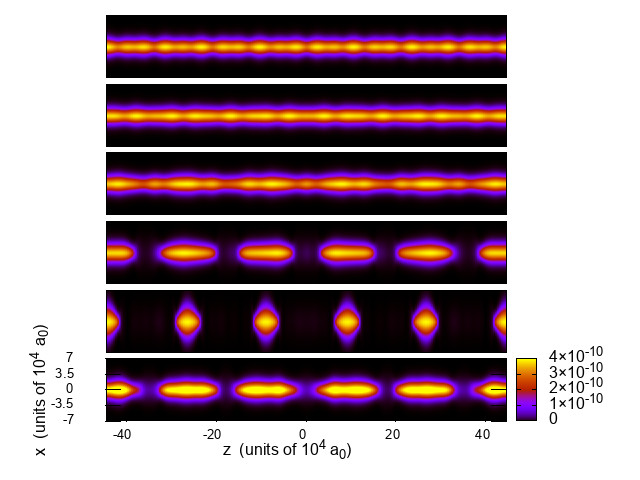}
}
\caption{
Snapshots from the real time evolution obtained by solving  the 1D equation for 
a free-standing filament with $N_1/L=0.159 \,a_0^{-1}$. 
The color bar displays the total number density in units of $a_0^{-3}$.
The filament is subject to an initial, most generic perturbation, as described in the text.
From top to bottom, the snapshots are taken at times $t=0,8,13,15,17$, and 19 ms.
}
\label{fig6}
\end{figure*}

The time  evolution obtained by starting the dynamics from this initial state is shown in the sequence 
of snapshots displayed in Fig. \ref{fig6}.
It appears that the filament undergoes fragmentation, 
leading to the appearance of five droplets. This fragmentation pattern is what
is expected from an undulation with wavelength $\lambda =L/5=1.6\times 10^5\,a_0$, where $L$ is the length of
the filament in the figure, which corresponds to a value $kR\sim 0.7$, close to the maximum
of the curve shown in Fig. \ref{fig3}.
Therefore, as expected, the fragmentation dynamics is eventually dominated by the fastest
mode compatible with the length of our simulation cell,
resulting in the filament fragmentation into 
regularly arranged, identical droplets. Once formed, they
execute a series of large amplitude oscillations, being alternately compressed and elongated in the
filament direction.
Eventually, these oscillations will be damped by any residual friction, resulting in a necklace of identical, equidistant quantum droplets.

\subsection{Possible supersolid behavior of the fragmented state}

From the previous results, the lowest energy state resulting from the 
fragmentation of a quantum liquid filament  
appears  to be made of regularly arranged
droplets, the inter-droplet spacing being determined by the 
wavelength of the fastest mode that matches the filament length $L$.

When the number of atoms of each species in the mixture is such that 
the local densities satisfy the equilibrium condition 
$N_1/N_2=\sqrt{g_{22}/g_{11}}$,
the resulting droplets are separated by vacuum.
However, when the population ratio deviates from the optimal value,
i.e., there is one species in excess,
then the extra atoms in the larger component (in the following the $^{87}$Rb species)
cannot bind to the droplets,
whose composition already satisfies the equilibrium ratio,
and form instead a uniform halo embedding them.
This  dilute superfluid background is expected to provide a degree
of global phase coherence to the system, unlike the case of droplets separated
by vacuum where no such coherence is present between adjacent droplets.
This suggests that the droplets array in the unbalanced system
may display a supersolid character, 
i.e., coexistence of superfluidity and a periodic density modulation.

The possibility of supersolid phases in a binary Bose mixture
has recently been put forward in Ref. \onlinecite{Sac20}, 
where a self-bound 2D
supersolid stripe-phase in a weakly interacting binary BEC with spin-orbit coupling
has been proposed,
being stabilized by the Lee-Huang-Yang beyond-mean-field term.
In Ref. \onlinecite{Ten22}, a single one-dimensional droplet made 
of a binary Bose mixture immersed into a background of the excess species
and subject to periodic boundary conditions (as a model for a droplet 
confined in a toroidal trap), was found to display 
non-classical rotational inertia and thus the coexistence
of rigid-body and superfluid character.

In our case, to verify the hypothesis that the fragmented state
in the presence of an excess population of atoms of one species
may indeed show supersolid character, we have looked for a
hallmark of supersolid behavior of the modulated
structures, namely a finite non-classical translational
inertia.
Following Refs. \onlinecite{Pom94,Sep08}, we define the superfluid fraction
$f_s$ as the fraction of particles that remain at rest in the
comoving frame with a constant velocity $v_x$
\begin{equation}
 f_s=1-\lim_{v_x\to 0} \frac{{\big <}(P_{x,1}+P_{x,2}){\big >}}{(N_1m_1+N_2m_2)v_x} \, ,
\end{equation}
where ${\big <}P_{x,i}{\big >}=-i\hbar \int \psi _i ^{\ast} \partial \psi _i/\partial x$
is the expectation value of the momentum of the $i$-th species and $(N_1m_1+N_2m_2)v_x $ is
the total momentum of the system as if all  droplets
were moving as a rigid body.
A non-zero value for $f_s$ reveals global phase
coherence in a periodic system like the one studied here.

The $f_s$ parameter should not be confused with the total superfluid fraction. For instance,
in the regime where self-bound droplets
form but are separated  from each other by vacuum, $f_s$ is zero
although droplets are individually superfluid,  
meaning that there is no global phase coherence.
At variance, for an ideal superfluid filament prior to breakup, 
$f_s$ is equal to 1 since the system is homogeneous along the filament axis.
Any periodic modulation should result in a finite value $0<f_s<1$.

\begin{figure}[!]
\centerline{
\includegraphics[width=1.0\linewidth,clip]{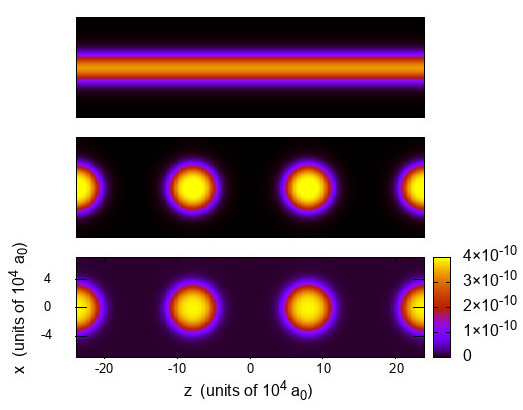}
}
\caption{Lowest energy configurations for the states described in the text.
Upper panel: free-standing filament prior to fragmentation;
middle panel: fragmented state (balanced mixture composition);
lower panel: fragmented state (unbalanced mixture composition).
The color bar displays the total number density in units of $a_0^{-3}$.
}
\label{fig7}
\end{figure}

\begin{figure}[!]
\centerline{
\includegraphics[width=1.0\linewidth,clip]{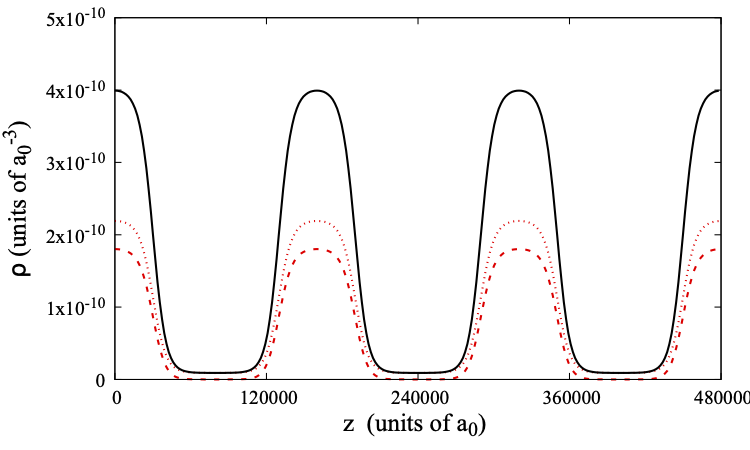}
}
\caption{Density profiles along the filament symmetry $z$-axis for the droplet array shown in the lower panel of Fig. \ref{fig7}.
Dotted line: $^{87}$Rb density; dashed line: $^{41}$K density; solid line: total ($^{41}$K+$^{87}$Rb) density.
}
\label{fig8}
\end{figure}

We have calculated $f_s$ for the three ground state structures shown in Fig. \ref{fig7}.
They refer to the case $N_1/L=0.159 \,a_0^{-1}$.
In the top panel the equilibrium filament is shown. The calculated value for the superfluid 
fraction is $f_s=1$, as expected.
The middle panel shows the fragmented, multi-droplet state resulting from a mixture satisfying the
equilibrium composition ratio when it has been subject to an initial perturbation with $kR\sim 0.7$.
We find for such structure $f_s< 0.001$.
The lowest panel shows instead the case where the $^{87}$Rb species is in excess.
The chosen values of $N_1$ and $N_2$ are such that there is just a small 
background density due to the excess species ($\sim 2.5\%$ of the total density in the center
of a droplet), as shown in 
Fig. \ref{fig8}, where a cut of the densities ($^{41}$K, $^{87}$Rb and total $^{41}$K+$^{87}$Rb densities)
along the system symmetry $z$-axis is displayed.
In spite of the small amount of $^{87}$Rb density enveloping the $^{41}$K-$^{87}$Rb droplets,
the resulting superfluid fraction is surprisingly large, $f_s=0.53$.
This finite value for $f_s$ likely indicates a supersolid character
of the droplet array.

\medskip

\section{Summary and outlook}

We predict that a bosonic $^{41}$K-$^{87}$Rb mixture confined in a sufficiently long 
toroidal trap, in the regime where a self-bound liquid state forms,
will undergo Rayleigh-Plateau instability and produce a necklace of droplets
inside the torus.
The droplets will be subject to oscillations also in the tubular confinement,
although some damping in the real system is expected what drives them to rest.

In the presence of a non-equilibrated mixture, i.e. when there is one species in excess
with respect to the optimal ratio $N_1/N_2=\sqrt{g_{22}/g_{11}}$, the excess species is expelled 
from the $^{41}$K-$^{87}$Rb filament. Due to the presence of transverse confinement,
the excess part cannot evaporate but remains instead in the
trap, enveloping the $^{41}$K-$^{87}$Rb liquid droplets resulting from filament fragmentation.
This results in a global phase coherence between one droplet and
the next, leading to a possible supersolid behavior.

This toroidal geometry is experimentally realizable and
therefore our results could be compared with experiments when 
curvature effects can be neglected. In practice, it might be 
very challenging to prevent the formation of a single big droplet
inside the toroidal trap during the quenching of $a_{12}$ required to 
reach the quantum liquid state. In this regard, it might help  to look for 
the instability when the value of the interspecies scattering length 
$a_{12}$ is close to the gas-liquid transition value so that the density of the
self-bound state is not much larger than that in the gas phase. In the presence
of a tight confinement, the single droplet state should be energetically disfavoured
with respect to the (unstable) filament configuration.

\begin{acknowledgments}
F.~A. acknowledges helpful discussions with Luca Salasnich.
This work has been  performed under Grant No.  PID2020-114626GB-I00 from the MICIN/AEI/10.13039/501100011033
and benefitted from COST Action CA21101 ``Confined molecular systems: 
form a new generation of materials to the stars'' (COSY) 
supported by COST (European Cooperation in Science and Technology).
\end{acknowledgments}

\end{document}